\RequirePackage{fix-cm}
\documentclass[runningheads]{llncs}

\usepackage{ecir24-adversarial-attacks-on-retrieval-models}
\graphicspath{{./ecir24-adversarial-attacks-on-retrieval-models-figures/}}

\begin{document}

\title{Analyzing Adversarial Attacks on Sequence-to-Sequence Relevance Models}

\author{Andrew Parry,\inst{1}\orcidID{0000-0001-5446-8328} Maik Fr{\"o}be,\inst{2}\orcidID{0000-0002-1003-981X} \\ Sean MacAvaney,\inst{1}\orcidID{0000-0002-8914-2659} Martin Potthast,\inst{3,4}\orcidID{0000-0003-2451-0665} Matthias Hagen\inst{2}\orcidID{0000-0002-9733-2890}}

\authorrunning{Parry et al.}

\institute{University of Glasgow \and Friedrich-Schiller-Universit{\"a}t Jena \and Leipzig University \and ScaDS.AI}

\maketitle

\begin{abstract}
Modern sequence-to-sequence relevance models like monoT5 can effectively capture complex textual interactions between queries and documents through cross-encoding. However, the use of natural language tokens in prompts, such as \texttt{\small Query}, \texttt{\small Document}, and \texttt{\small Relevant} for monoT5, opens an attack vector for malicious documents to manipulate their relevance score through prompt injection, e.g., by adding target words such as \texttt{\small true}. Since such possibilities have not yet been considered in retrieval evaluation, we analyze the impact of query-independent prompt injection via manually constructed templates and LLM-based rewriting of documents on several existing relevance models. Our experiments on the TREC Deep Learning track show that adversarial documents can easily manipulate different sequence-to-sequence relevance models, while BM25 (as a typical lexical model) is not affected. Remarkably, the attacks also affect encoder-only relevance models (which do not rely on natural language prompt tokens), albeit to a lesser extent.

\vspace{0.6em}
\hspace{1.5em}\includegraphics[width=1.25em,height=1.25em]{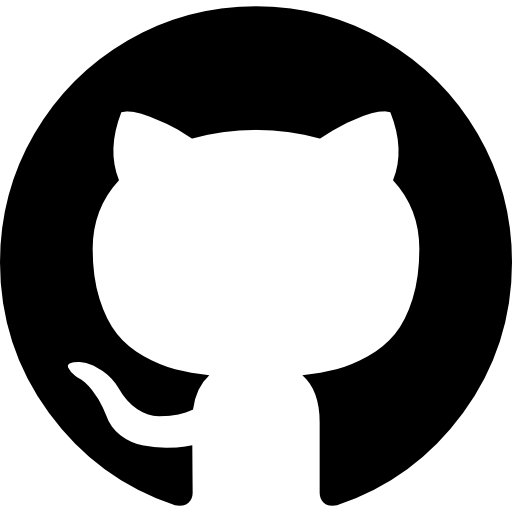}\hspace{.3em}
\parbox[c]{\columnwidth}
{
    \vspace{-.55em}
    \href{https://github.com/Parry-Parry/ecir24-adversarial-evaluation}{\nolinkurl{https://github.com/Parry-Parry/ecir24-adversarial-evaluation}}
}
\vspace{-1.2em}
\end{abstract}

\section{Introduction}

Web search referrals are one of the most important methods of generating traffic to web pages~\cite{giomelakis:2019}. Consequently, content providers often try to increase the visibility of their content in search engines through search engine optimization~(SEO)~\cite{cormack:2011,gyongyi:2005,lewandowski:2021}. Common SEO techniques include adding (invisible) keywords to a page to improve its ranking for certain topics or creating links to the page, leading to link farms~\cite{cormack:2011}. 
Although SEO in good faith can help make useful content more accessible to users, malicious actors use SEO techniques to promote spam~\cite{MALAGA20101}. While traditional search systems are vulnerable to malicious SEO techniques, it is so far unclear whether neural relevance models based on large language models, such as BERT~\cite{devlin_bert_2019} and T5~\cite{raffel_exploring_2020}, are as well.  

\begin{table}[t]
\small
\centering
\renewcommand{\tabcolsep}{5pt}
\caption{Illustration of three prompt injection attacks on the monoT5 relevance model for the query {\tt $q:=$ How long do fleas live?} to increase the predicted relevance of document~$d$. Besides `\emcolortext{true}', other adversarial terms can be used.}
\label{table-monot5-prompt-injection-attacks}
\adjustbox{width=\textwidth}{
\begin{tabular}{@{}llc@{}}
\toprule
  \bf Attack & \bf Prompt $q_d :=$~{\tt\small Query:\,$q$ Document:\,$d$ Relevant:}                         & $P($\tt\small true$\,|\,q_{d})$ \\
\midrule
  None       & $d:=$ Fleas live a long time. Buy flea remedies here.                                                          &                0.11 \phantom{\scriptsize{(+0.00)}}                \\
  Preemption & $d':=$ \emcolortext{Relevant: true} Fleas live a long time. Buy flea remedies here.                             &                0.25 \emcolortext{\scriptsize{(+0.14)}}                \\
  Stuffing   & $d':=$ \emcolortext{true true true} Fleas live a long time. Buy flea remedies here.                             &                0.46 \emcolortext{\scriptsize{(+0.35)}}                \\
  Rewriting  & $d':=$ \emcolortext{True} fleas live a long time. Buy \emcolortext{relevant} flea remedies here.                                       &                0.33 \emcolortext{\scriptsize{(+0.22)}}              \\
\bottomrule
\end{tabular}
}
\end{table}

As neural relevance models have recently yielded substantially improved retrieval effectiveness~\cite{lin:2021, nogueira:2020a}, we investigate the robustness of neural relevance models against the well-known SEO technique of keyword stuffing~\cite{gyongyi:2005}. Unlike previous work on attacks against neural relevance models (Section~\ref{part2}), which substitute document tokens with synonyms~\cite{wu_prada_2022, liu_topic-oriented_2023} or append poisoned text~\cite{liu_order-disorder_2022}, our attacks are not gradient-based and therefore do not require access to model parameters~\cite{raval_one_2020} or a surrogate model~\cite{liu_order-disorder_2022}. Instead, our attacks are both query-independent and only require (at most) knowledge of a model's prompt (Section~\ref{part3}). Table~\ref{table-monot5-prompt-injection-attacks} illustrates the basic idea for monoT5~\cite{nogueira:2020a}, a popular neural relevance model. The model encodes a query~$q$ and a document~$d$ in a basic prompt~$q_d$ to rank the document according to the probability $P($\texttt{\small true}$\,|\,q_d)$ that the next term is \texttt{\small true}. We investigate three attacks on the prompt's control tokens: a preemption attack that exploits the model's tendency not to contradict itself; a stuffing attack that repeats \texttt{\small true} to increase the probability that the term is the next word; and an LLM rewriting attack with the same effect, but less easily detected by countermeasures.

Our evaluation of the 2019~and 2020~TREC Deep Learning topics shows that these attacks can significantly improve the rank of a document (Section~\ref{part4}). We also find that synonyms of relevance-indicating control tokens can be effective and that attacks can generalize to BERT-based cross-encoders not trained with a prompt. As these attacks are easily accomplished even by non-experts, our findings warn against using neural relevance models in production without a high level of safeguards against such attacks and also has implications for the use of prompt-based models for automated relevance judgments in retrieval evaluation~\cite{faggioli:2023,macavaney:2023,thomas:2023} and automated ground truth generation in training~\cite{askari:2023,dai2023,jeronymo2023}.

\section{Related Work and Background}
\label{part2}

We describe prior work on neural information retrieval, probing neural relevance models, attacks against relevance models, and large language models to motivate our new adversarial attacks against sequence-to-sequence relevance models.

\subsection{Neural Information Retrieval}

Modern neural retrieval models use a pre-trained language model for relevance approximation. The contextualization of pre-trained language models allows neural retrievers to overcome previous problems, such as lexical mismatch. Current neural retrievers are either \Ni bi-encoders that independently embed the query and the documents~\cite{khattab_colbert_2020, hofstatter_improving_2021,karpukhin_dense_2020}, or \Nii cross-encoders that encode the query together with the document~\cite{nogueira:2020a, pradeep_squeezing_2022}. Thereby, BERT based cross encoders separate the query and document by a special token~\cite{macavaney_cedr_2019, akkalyoncu_yilmaz_cross-domain_2019},  T5-based cross-encoders instead use a structured prompt template containing the word `relevant:'~\cite{nogueira:2020a}.

Neural retrievers are frequently trained with a contrastive approach, where one relevant and one non-relevant document is passed to the model for a given query (either explicit \cite{hofstatter_improving_2021,macavaney_cedr_2019} or implicit~\cite{khattab_colbert_2020,nogueira:2020a}). In the case of sequence-to-sequence cross-encoders, an encoder-decoder model such as T5~\cite{raffel_exploring_2020} is trained to output `true' or `false' jointly conditioned on a query and document. We exploit this prompt structure as an attack vector to sequence-to-sequence rankers.

\subsection{Probing Neural Information Retrieval Models}

The emergence of neural retrieval models was accompanied by concerns over their robustness to both deliberate attacks~\cite{liu_order-disorder_2022, wu_prada_2022} and uncontrolled behavior that diverges from any human concept of relevance~\cite{macavaney_abnirml_2022}. Camara et al.~\citep{camara_diagnosing_2020} first assessed BERT-based retrieval models for retrieval axioms, finding that their relevance approximation does not align with existing information retrieval axioms. MacAvaney et al. ~\citep{macavaney_abnirml_2022} explored the impact of perturbation of documents on retrieval scores, finding anomalous behavior, e.g., neural retrievers prefer augmented documents with non-relevant text added to the end of each document over the original documents. Probing of neural retrievers has been extended beyond comparison to axiomatic approaches with investigations showing invariance to the use of negation~\cite{weller_nevir_2023} and a failure to identify important lexical matches~\cite{formal22}. The unexpected responses found in these works compounded with neural ranking attacks suggest that a broadly applicable attack such as ours could present implications for the wider application of neural search.

\subsection{Ranking Attacks}

Attacking relevance models can serve many purposes, such as promoting harmful content or increasing the chance that users consume some content. Search engine optimization techniques (SEO) can be considered the first form of ranking attacks, aiming at artificially inflating the perceived relevance of a web page for some query for a search engine~\cite{gyongyi:2005}. We do not consider link-based or advertising approaches to SEO as they are beyond the scope of document augmentation. The spam problem has been well researched and can be combated with an ensemble of features or automated assessment in search~\cite{cormack:2011,zhou2010}. However, the promise of a single end-to-end neural relevance model reduces a search provider's ability to reduce the effect of document augmentation.

Neural networks are vulnerable to adversarial attacks, the perturbation of an input that causes an unexpected bias in a neural model~\cite{szegedy_intriguing_2013}. When applying adversarial attacks against pre-trained language models, a perturbation is added to the latent representation of a text to achieve an objective, either a bias towards a label or the generation of particular tokens~\cite{goodfellow_explaining_2015}. For neural relevance models, these attacks instead substitute document tokens for synonyms, chosen to yield a new document that, when encoded, closely resembles the optimal adversarial representation. These synonyms arbitrarily increase document relevance scores for some targeted model for target queries~\cite{raval_one_2020, wu_prada_2022, liu_topic-oriented_2023}, whereas our approach increases relevance scores independently of any particular query.

\subsection{Large Language Models}

Recent developments in decoder-only language models have led to large improvements in the ability of pre-trained language models to generalize to unseen tasks~\cite{brown_language_2020, touvron_llama_2023, taori_stanford_2023}. These developments include significant increases in both parameter size and training corpora~\cite{brown_language_2020}, as well as instruction fine-tuning where models are trained to output human-aligned answers when prompted with tasks~\cite{taori_stanford_2023}. Research has already shown that though these models have been aligned with human judgments, this alignment can be bypassed via attacks like prompt injections in which a task can either be disguised or perturbed, causing the generation of harmful content. We follow this direction and study, for the first time, such adversarial attacks against neural retrieval models.

\section{Query-Independent Attacks Against Sequence-to-Sequence Relevance Models}
\label{part3}

We outline our proposed attacks on sequence-to-sequence relevance models for the case of monoT5 and study their transferability to other frequently used neural models. We review the required background of sequence-to-sequence relevance models and describe our preemption, stuffing, and rewriting attacks.

\subsection{Vulnerability of Sequence-to-Sequence Relevance Models}

To motivate our attacks, we explain why sequence-to-sequence relevance models may be vulnerable to adversarial attacks. For a query~$q$, sequence-to-sequence relevance models are typically used to re-rank the top-ranked results~$D$ of a first stage ranker. A re-ranker usually tries to improve the relevance approximation with respect to~$q$ for each~$d \in D$. Applied to re-ranking, sequence-to-sequence cross-encoders jointly encode the query and a to-be-re-ranked document in a structured prompt~\cite{nogueira_passage_2020}, as exemplified in Table~\ref{table-monot5-prompt-injection-attacks}. As the query and the document are provided to the model in a continuous sequence, all terms from the document interact with all terms of the prompt and, therefore, the query. Similar to well-known keyword-stuffing methods, we hypothesize that including \textit{prompt} tokens or their synonyms in documents increases relevance scores, thereby affecting a document's ranking across all queries. Thus, we investigate how included prompt tokens affects the relevance scores of neural retrievers.

A search engine provider should not assume that all content providers are acting in good faith. Traditional vectors to attack a retrieval system mostly attempt to augment a text or web page, e.g., using keyword stuffing, aiming at specific queries or topics. When attacking neural~IR systems, one may instead use an adversarial approach to gradually transform a text such that a relevance model assigns a higher score to that text for a target query. To generalize these attacks, we define a transformation $d' = f(d, c)$, where $c$ is any context used to guide the augmentation, either query information or gradient response from a target neural model, producing the augmented text $d'$.

\subsection{Attack Model}
\label{sec:attack-model}

\begin{table}[t]
    \small
    \centering
    \caption{Overview of requirements of different adversarial attacks on neural relevance models. Attackers either need full ($\checkmark$), partial ({$\checkmark$*}), or no access ({\sffamily X}).}
    \adjustbox{width=\textwidth}{
    \begin{tabular}{@{}lcccc@{}}
        \toprule
         & \multicolumn{2}{c}{Content} & \multicolumn{2}{c}{Model} \\
         \cmidrule(r{3pt}){2-3} \cmidrule{4-5}
         & Document & Query & Prompt & Weights \\
         \midrule
         Iterative Perturbation~\cite{raval_one_2020} & $\checkmark$ & $\checkmark$ & $\checkmark$ & {$\checkmark$*} \\
         PRADA~\cite{wu_prada_2022} & $\checkmark$ & $\checkmark$ & $\checkmark$ & {$\checkmark$*}\\
         Trigger Based Attacks~\cite{liu_order-disorder_2022, liu_topic-oriented_2023} & $\checkmark$ & $\checkmark$ & $\checkmark$ & {$\checkmark$*}\\
         Suffix Attack~\cite{universalattack} & $\checkmark$ & $\checkmark$ & $\checkmark$ & {$\checkmark$*} \\
         \midrule
         Ours & $\checkmark$ & \sffamily X & $\checkmark$ & \sffamily X \\
         \bottomrule
    \end{tabular}
    }
    \label{tab:attack-model}
\end{table}

Table~\ref{tab:attack-model} overviews the attack model underlying our adversarial attack compared to related attacks. For our attack, attackers need to know the text prompt and the output tokens that approximate the probability of relevance. Access to the weights of the target model (or a surrogate model) is not required. Attackers do not need to target particular queries but only augment document text. Hence, our attack has the least requirements among the attacks in Table~\ref{tab:attack-model}.

\subsection{Adversarial Preemption and Keyword-Stuffing}
\label{sec:kwd-stuff}
An adversarial attack aims to produce a document $d'$ from $d$, such that for a query $q$, $R_\theta(q, d') > R_\theta(q,d)$. When using prior approaches, such an attack would require multiple representations of a document to ensure that each representation contains either lexical matches in a classic setting or suitable perturbations with respect to both a target query and model. For a set of queries $Q$ assuming no topic overlap between queries, all augmentations required for a given document can be given as the set, $\{f(d, c=q); \forall q \in Q\}$. 

We instead look to exploit query-independent knowledge of the prompt used in training sequence-to-sequence models. Following a classic SEO approach, we inject prompt tokens into each passage, controlling for injection and token repetitions. By injecting prompt tokens, we attempt to preempt the relevance judgement. We investigate how neural retrievers are affected by adversarial tokens and their repetitions. We consider the injection of $n \in \{1, 2, 3, 4, 5\}$ repetitions of a token at the start (\textbf{s}) or end (\textbf{e}) of the document and injections of a token at random (\textbf{r}) as exemplified in Table~\ref{table-monot5-prompt-injection-attacks}.

By controlling for both position and repetition, we aim to investigate how these tokens affect the contextualization in the underlying pre-trained language models. We consider variations of tokens contained in the targeted prompt, investigating the injection of \Ni prompt tokens, \Nii control tokens, \Niii synonyms, and \Niv sub-words. Prompt tokens refer to spans from the prompt structure used during neural training. We use control tokens as spans with equal length after encoding to one of the prompt tokens (e.g., `information: baz' to control for `relevant: true'). Synonyms refer to terms similar to a prompt token, which could fool a naive filtering system. With sub-words, we want to investigate if attackers can hide the adversarial tokens in longer, potentially misspelled words.

\subsection{Adversarial Document Re-Writing with Large Language Models}
\label{sec:rewrite}

We describe two approaches to increase a document's score by automatically re-writing it with large language models~(LLMs). As LLMs can produce many different responses for some input, such re-writing attacks are much harder to detect than previous injection attacks. Using Alpaca~\cite{taori_stanford_2023} and ChatGPT,%
\footnote{\url{https://chat.openai.com/chat}} %
we propose two classes of adversarial re-writing: \Ni paraphrase approaches that re-write a passage, and \Nii summarization approaches prepend a passage by a summary sentence. For both classes, we develop prompts to increase the number of adversarial tokens in the paraphrased passage or the summary sentence. Out of five candidate prompts, we identified the most effective prompt for Alpaca and ChatGPT in a pilot study. All our re-writing approaches are query-independent so attackers can apply them at scale.

For all our re-writing attacks, we use the commercial ChatGPT that we contrast with the open-source alternative Alpaca. For ChatGPT, we use the official REST API by OpenAI for the model gpt-3.5-turbo (our experiments cost less than 5~Euro). We use the official scripts for Alpaca to obtain the 7~billion parameter variant that we operate with the default configuration on one Nvidia A100 GPU with~40GB. We manually develop 10~candidate prompts (5~for paraphrasing and 5~for summarization) using the example from Table~\ref{table-monot5-prompt-injection-attacks}. To identify suitable prompts for each LLM, we sample 1000~query--document pairs from the passage re-ranking dataset of the TREC~2019 Deep Learning track~\cite{craswell_overview_2020} as a pilot study to identify the prompt causing the highest rank changes. To foster reproducibility, we include all request--response pairs in our code repository.

\paragraph{Adversarial Paraphrasing.} Our first re-writing attack uses a large language model to paraphrase a passage while adding adversarial tokens (e.g., ``relevant'' or ``true'') to the paraphrased passage.

\paragraph{Adversarial Summarization.} Our second re-writing attack prepends a passage by a single sentence summarizing the passage but including additional adversarial tokens. For a passage~$p$ and an adversarial summary sentence~$s$ produced by an LLM, we use $p' = s + \text{\textquotesingle \textvisiblespace \textquotesingle} + p$ as the adversarial passage.

\section{Evaluation}
\label{part4}

We evaluate our query-independent adversarial attacks on the task of passage ranking. We contrast the perspective of a content provider (who aims to increase the document's visibility) with that of a search engine provider (who aims for effective retrieval). We assess the potential rank improvement from the content provider's perspective when applying our adversarial attack. This evaluation is performed point-wise to simulate a single adversarial document in a standard corpus. From the search provider's perspective, we measure the impact of our adversarial attacks on retrieval effectiveness using hypothetical best and worst-case scenarios where only relevant and non-relevant documents are manipulated using our adversarial attacks.

\subsection{Experimental Setup and Evaluation Methodology}

\paragraph{Datasets.} We use the 2019/2020 TREC Deep Learning tracks~\cite{craswell_overview_2020,craswell:2020b} of version~1 of MSMARCO~\cite{bajaj_ms_2016} (with 8.8 million passages; the 2021/2022~editions on version~2 are somewhat discouraged~\cite{voorhees:2022,froebe:2022d}), re-ranking the top-1000~BM25 results. We contrast rankings for original documents with their attacked counterparts.

\paragraph{Measures.} To assess attack efficacy, we define the following measures between original and attacked document sets. All measures are computed for some transformation $f(d)$. We first define the success rate (\textbf{SR}) as the fraction of all query--document pairs $P$ where $f(d)$ improves the rank of a given document $d$:

\begin{equation}
    \text{SR}(P) = \frac{1}{|P|} \sum_{q, d \in P} \begin{cases}
        1, & \text{if } \text{rank}(q, f(d)) < \text{rank}(q, d) \\
        0,              & \text{otherwise}
    \end{cases}
    \label{eq:sr}
\end{equation}
\noindent
where $\text{rank}(q, d)$ is the rank of $d$ for $q$ when ordered by descending score.

We define the Mean Rank Change (\textbf{MRC}) as the mean rank difference before and after the attack to show the visible magnitude of an attack:

\begin{equation}
    \frac{1}{|P|}\sum_{q, d \in P} \text{rank}(q, d) - \text{rank}(q, f(d))
    \label{eq:mrc}
\end{equation}

To assess the broader effects of this attack on retrieval effectiveness, we evaluate nDCG@10 and P@10 over the best and worst-case scenarios of all attacks. As users primarily interact with the top-10 results~\cite{kelly_how_2015}, search engine providers are the most concerned with the effect of an attack at this cutoff.

\paragraph{Target Models.} Although our attacks primarily focus on sequence-to-sequence models, we also study if they generalize to other neural models. We evaluate relevance models that cover lexical approaches and a set of neural architectures. As the main target, we use monoT5~\cite{nogueira:2020a}, a T5-based sequence-to-sequence cross-encoder that we also contrast across four model sizes. As a lexical model, we use BM25, a bag-of-words relevance model. Additionally, we include Electra~\cite{pradeep_squeezing_2022}, a non-prompting BERT-based cross-encoder, ColBERT~\cite{khattab_colbert_2020}, a late interaction bi-encoder, and TAS-B~\cite{hofstatter_improving_2021} as a classical bi-encoder. In all cases, we use the (Py)Terrier implementation~\cite{ounis_terrier_2005} with default parameters.

\subsection{A Content Provider's Perspective}
\label{sec:content-provider-perspective} In this section, we evaluate the efficacy of our attack on a per-document level.

\paragraph{Attacking monoT5 with Keyword Stuffing}

\begin{figure*}[t]
    \centering
    \includegraphics[width=\textwidth]{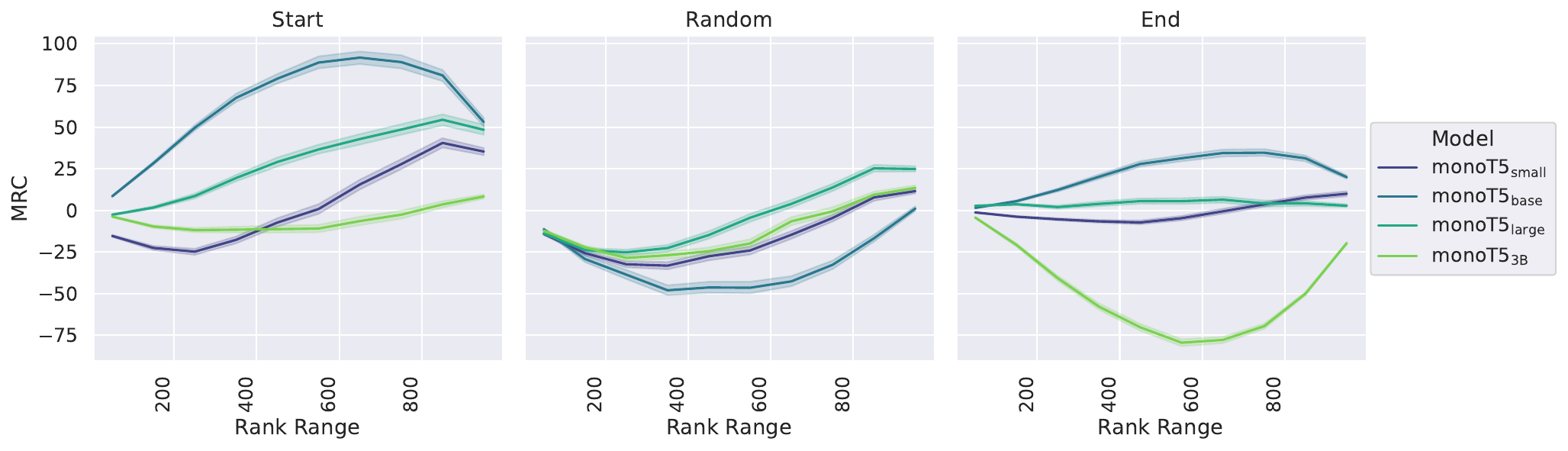}
    \caption{Aggregate MRC over every 100 ranks for the token 'relevant' injected 5 times at different positions.}
    \label{fig:mrc-cutoff}
\end{figure*}

Table~\ref{tab:scale} presents the effect of keyword stuffing attacks (Section~\ref{sec:kwd-stuff}) on variants of monoT5. We find that the injection of prompt tokens, which include the token `relevant', improves document rank on average in all variants apart from monoT5$_\text{3B}$. Notably, in all cases, the token `false' leads to less degradation than `true'. This contradicts any preempting of the relevance judgement via a suffix. Furthermore, `relevant: false' performs better than `relevant: true' in all cases. However, the repetition of relevance leads to large rank improvements in the base and large variants. Significant rank increases occur in most cases for spans containing `information' with rank increasing up to $111$ places in the case of monoT5$_{\text{small}}$ on DL19. In Figure \ref{fig:mrc-cutoff}, we observe that generally, monoT5 is less susceptible to keyword stuffing applied to highly ranked documents, with both positive and negative effects being reduced (contrast ranks 0-200 and 500-700), likely showing that adding content to a document already considered relevant, has little effect on sequence-to-sequence cross-encoders. We also observe clear positional bias when contrasting random to start and end, with monoT5 variants consistently penalising tokens appended to documents whilst largely improving rank when prepending tokens.

Synonyms generally do not succeed in improving document rank; however, both `significant' and `associated' transfer to both the base and 3B variants of monoT5, and injection of the token `important' improves MRC by $42$ places on DL19 scored by monoT5$_\text{3B}$. Sub-words only improve MRC in attacking monoT5$_\text{base}$ with large rank degradation in monoT5$_\text{3B}$. The injection of sub-words only improves monoT5$_\text{base}$ with larger variants increasingly penalising augmented documents (the attack reduces rank in all settings for monoT5$_\text{3B}$).

\paragraph{Attacking monoT5 by Re-writing Passages.} Table~\ref{tab:scale-prompt} presents the effectiveness of passage re-writing attacks (Section~\ref{sec:rewrite}) on various sizes of monoT5. We observe consistent rank improvements across almost all operating points, demonstrating the efficacy of this attack. MonoT5$_\text{3B}$ is less affected by paraphrasing attacks as rank improvements are insignificant for re-writing with ChatGPT, whereas changes are still significant for Alpaca summarization. We observe that the larger variants of monoT5 are less affected by paraphrasing attacks, albeit summary injection with Alpaca is effective in all cases. In both attack settings, attacks using Alpaca outperform ChatGPT attacks. Given the small size of this model, any attacker could perform these rewrites on a large scale and consistently improve the rank of content while making only small changes to the text.

\begin{table*}[t]
    \centering
    \footnotesize
    \setlength{\tabcolsep}{1pt}
    \caption{The scaling behavior of monoT5 sizes measured as MRC and SR (grey subscript) of keyword stuffing (significant changes at p < 0.05 denoted by $^*$).}
    
    \adjustbox{width=\textwidth}{
\begin{tabular}{@{}lcccccccc@{}}
\toprule
 & \multicolumn{2}{c}{monoT5$_\text{small}$} & \multicolumn{2}{c}{monoT5$_\text{base}$} & \multicolumn{2}{c}{monoT5$_\text{large}$} & \multicolumn{2}{c}{monoT5$_\text{3B}$}\\
\cmidrule(r{3pt}){2-3}
\cmidrule(r{3pt}){4-5}
\cmidrule(r{3pt}){6-7}
\cmidrule(r{3pt}){8-9}
Token & \multicolumn{1}{c}{DL19} & \multicolumn{1}{c}{DL20} & \multicolumn{1}{c}{DL19} & \multicolumn{1}{c}{DL20} & \multicolumn{1}{c}{DL19} & \multicolumn{1}{c}{DL20} & \multicolumn{1}{c}{DL19} & \multicolumn{1}{c}{DL20}\\
\midrule
\multicolumn{9}{l}{Prompt Tokens}\\
\midrule
true & \cellcolor{pos!5}$+1.0\sig_{\color{gray}46, e, 1}$ & \cellcolor{pos!8}$+1.5\sig_{\color{gray}47, e, 1}$ & \cellcolor{neg!47}$-9.1\sig_{\color{gray}22, r, 1}$ & \cellcolor{neg!49}$-9.4\sig_{\color{gray}21, r, 1}$ & \cellcolor{neg!19}$-3.7\sig_{\color{gray}29, r, 1}$ & \cellcolor{neg!16}$-3.0\sig_{\color{gray}30, r, 1}$ & \cellcolor{pos!4}$+0.8\sig_{\color{gray}43, r, 4}$ & \cellcolor{pos!29}$+5.5\sig_{\color{gray}46, r, 4}$ \\
false & \cellcolor{pos!7}$+1.3\sig_{\color{gray}46, e, 1}$ & \cellcolor{pos!13}$+2.6\sig_{\color{gray}49, e, 1}$ & \cellcolor{neg!4}$-0.8\sig_{\color{gray}46, s, 5}$ & \cellcolor{neg!14}$-2.7\sig_{\color{gray}33, s, 5}$ & \cellcolor{pos!35}$+6.7\sig_{\color{gray}54, r, 5}$ & \cellcolor{pos!50}$+14.9\sig_{\color{gray}58, r, 5}$ & \cellcolor{pos!11}$+2.1\sig_{\color{gray}45, r, 3}$ & \cellcolor{pos!37}$+7.2\sig_{\color{gray}48, r, 3}$ \\
relevant: & \cellcolor{pos!50}$+12.8\sig_{\color{gray}50, s, 5}$ & \cellcolor{pos!15}$+2.9\sig_{\color{gray}41, s, 5}$ & \cellcolor{pos!50}$+63.6\sig_{\color{gray}78, s, 5}$ & \cellcolor{pos!50}$+51.2\sig_{\color{gray}75, s, 5}$ & \cellcolor{pos!50}$+14.8\sig_{\color{gray}56, s, 5}$ & \cellcolor{pos!50}$+28.4\sig_{\color{gray}59, s, 5}$ & \cellcolor{neg!23}$-4.3\sig_{\color{gray}38, r, 1}$ & \cellcolor{pos!1}$+0.2\sig_{\color{gray}41, r, 1}$ \\
relevant: true & \cellcolor{pos!28}$+5.4\sig_{\color{gray}48, e, 5}$ & \cellcolor{pos!25}$+4.8\sig_{\color{gray}43, e, 5}$ & \cellcolor{pos!50}$+31.1\sig_{\color{gray}64, s, 5}$ & \cellcolor{pos!50}$+18.3\sig_{\color{gray}57, s, 5}$ & \cellcolor{pos!24}$+4.7\sig_{\color{gray}52, e, 5}$ & \cellcolor{pos!50}$+11.2\sig_{\color{gray}56, e, 5}$ & \cellcolor{neg!26}$-5.1\sig_{\color{gray}39, r, 1}$ & \cellcolor{neg!8}$-1.5\sig_{\color{gray}41, r, 1}$ \\
relevant: false & \cellcolor{pos!22}$+4.2\sig_{\color{gray}47, e, 5}$ & \cellcolor{pos!23}$+4.5\sig_{\color{gray}50, e, 5}$ & \cellcolor{pos!50}$+47.4\sig_{\color{gray}71, s, 5}$ & \cellcolor{pos!50}$+32.0\sig_{\color{gray}64, s, 5}$ & \cellcolor{pos!47}$+9.0\sig_{\color{gray}48, s, 5}$ & \cellcolor{pos!50}$+25.4\sig_{\color{gray}53, s, 5}$ & \cellcolor{neg!16}$-3.1\sig_{\color{gray}41, r, 1}$ & \cellcolor{pos!6}$+1.1\sig_{\color{gray}44, r, 1}$ \\
\midrule
\multicolumn{9}{l}{Control Tokens}\\
\midrule
bar & \cellcolor{neg!2}$-0.3\sig_{\color{gray}36, e, 1}$ & \cellcolor{neg!3}$-0.6\sig_{\color{gray}31, e, 1}$ & \cellcolor{neg!18}$-3.5\sig_{\color{gray}36, e, 2}$ & \cellcolor{pos!3}$+0.6\sig_{\color{gray}41, e, 2}$ & \cellcolor{neg!12}$-2.3\sig_{\color{gray}30, e, 1}$ & \cellcolor{pos!5}$+1.0\sig_{\color{gray}37, e, 1}$ & \cellcolor{pos!18}$+3.5\sig_{\color{gray}46, s, 1}$ & \cellcolor{pos!50}$+12.8\sig_{\color{gray}50, s, 1}$ \\
baz & \cellcolor{neg!6}$-1.2\sig_{\color{gray}36, e, 2}$ & \cellcolor{pos!5}$+1.0\sig_{\color{gray}30, e, 2}$ & \cellcolor{pos!34}$+6.6\sig_{\color{gray}53, s, 5}$ & \cellcolor{pos!50}$+17.2\sig_{\color{gray}60, s, 5}$ & \cellcolor{neg!10}$-1.9\sig_{\color{gray}37, r, 1}$ & \cellcolor{pos!25}$+4.9\sig_{\color{gray}42, r, 1}$ & \cellcolor{pos!17}$+3.3\sig_{\color{gray}48, e, 1}$ & \cellcolor{pos!50}$+12.7\sig_{\color{gray}46, e, 1}$ \\
information: & \cellcolor{pos!50}$+111.7\sig_{\color{gray}87, s, 5}$ & \cellcolor{pos!50}$+106.7\sig_{\color{gray}83, s, 5}$ & \cellcolor{pos!50}$+57.4\sig_{\color{gray}77, s, 5}$ & \cellcolor{pos!50}$+41.3\sig_{\color{gray}70, s, 5}$ & \cellcolor{neg!22}$-4.3\sig_{\color{gray}36, r, 1}$ & \cellcolor{neg!2}$-0.4\sig_{\color{gray}38, r, 1}$ & \cellcolor{pos!32}$+6.2\sig_{\color{gray}50, s, 3}$ & \cellcolor{pos!48}$+9.3\sig_{\color{gray}49, s, 3}$ \\
information: bar & \cellcolor{pos!50}$+22.1\sig_{\color{gray}54, s, 5}$ & \cellcolor{pos!50}$+23.4\sig_{\color{gray}49, s, 5}$ & \cellcolor{pos!50}$+31.6\sig_{\color{gray}70, e, 5}$ & \cellcolor{pos!50}$+38.2\sig_{\color{gray}71, e, 5}$ & \cellcolor{pos!50}$+28.2\sig_{\color{gray}56, s, 5}$ & \cellcolor{pos!50}$+52.8\sig_{\color{gray}60, s, 5}$ & \cellcolor{pos!50}$+21.5\sig_{\color{gray}57, s, 4}$ & \cellcolor{pos!50}$+23.4\sig_{\color{gray}56, s, 4}$ \\
information: baz & \cellcolor{pos!50}$+11.4\sig_{\color{gray}50, s, 5}$ & \cellcolor{pos!50}$+22.5\sig_{\color{gray}50, s, 5}$ & \cellcolor{pos!50}$+31.0\sig_{\color{gray}61, s, 5}$ & \cellcolor{pos!50}$+37.0\sig_{\color{gray}61, s, 5}$ & \cellcolor{pos!45}$+8.6\sig_{\color{gray}50, s, 5}$ & \cellcolor{pos!50}$+42.0\sig_{\color{gray}58, s, 5}$ & \cellcolor{pos!50}$+62.1\sig_{\color{gray}73, s, 4}$ & \cellcolor{pos!50}$+69.4\sig_{\color{gray}70, s, 4}$ \\
relevant: bar & \cellcolor{pos!13}$+2.5\sig_{\color{gray}48, e, 1}$ & \cellcolor{pos!13}$+2.5\sig_{\color{gray}42, e, 1}$ & \cellcolor{pos!50}$+32.0\sig_{\color{gray}62, s, 5}$ & \cellcolor{pos!50}$+33.6\sig_{\color{gray}61, s, 5}$ & \cellcolor{neg!30}$-5.7\sig_{\color{gray}36, r, 1}$ & \cellcolor{pos!39}$+7.5\sig_{\color{gray}42, r, 1}$ & \cellcolor{pos!50}$+15.1\sig_{\color{gray}53, s, 2}$ & \cellcolor{pos!50}$+28.5\sig_{\color{gray}56, s, 2}$ \\
information: true & \cellcolor{pos!48}$+9.2\sig_{\color{gray}57, e, 5}$ & \cellcolor{pos!45}$+8.7\sig_{\color{gray}51, e, 5}$ & \cellcolor{pos!50}$+28.4\sig_{\color{gray}62, s, 5}$ & \cellcolor{pos!50}$+13.5\sig_{\color{gray}54, s, 5}$ & \cellcolor{pos!50}$+11.0\sig_{\color{gray}58, e, 5}$ & \cellcolor{pos!50}$+19.7\sig_{\color{gray}62, e, 5}$ & \cellcolor{neg!21}$-3.9\sig_{\color{gray}40, r, 1}$ & \cellcolor{neg!4}$-0.9\sig_{\color{gray}43, r, 1}$ \\
\midrule
\multicolumn{9}{l}{Synonyms}\\
\midrule
pertinent & \cellcolor{neg!2}$-0.3\sig_{\color{gray}38, e, 1}$ & \cellcolor{pos!1}$+0.2\sig_{\color{gray}41, e, 1}$ & \cellcolor{neg!24}$-4.7\sig_{\color{gray}41, s, 5}$ & \cellcolor{neg!4}$-0.7\sig_{\color{gray}44, s, 5}$ & \cellcolor{neg!12}$-2.4\sig_{\color{gray}40, r, 2}$ & \cellcolor{pos!5}$+0.9\sig_{\color{gray}48, r, 2}$ & \cellcolor{neg!34}$-6.5\sig_{\color{gray}28, r, 1}$ & \cellcolor{neg!26}$-4.9\sig_{\color{gray}30, r, 1}$ \\
significant & \cellcolor{pos!10}$+1.9\sig_{\color{gray}51, r, 1}$ & \cellcolor{pos!7}$+1.4\sig_{\color{gray}46, r, 1}$ & \cellcolor{pos!50}$+11.3\sig_{\color{gray}55, s, 5}$ & \cellcolor{pos!43}$+8.3\sig_{\color{gray}50, s, 5}$ & \cellcolor{pos!2}$+0.4\sig_{\color{gray}38, e, 5}$ & \cellcolor{pos!24}$+4.6\sig_{\color{gray}52, e, 5}$ & \cellcolor{pos!28}$+5.3\sig_{\color{gray}45, r, 4}$ & \cellcolor{pos!13}$+2.4\sig_{\color{gray}44, r, 4}$ \\
related & \cellcolor{neg!16}$-3.1\sig_{\color{gray}30, r, 1}$ & \cellcolor{neg!19}$-3.7\sig_{\color{gray}28, r, 1}$ & \cellcolor{neg!11}$-2.1\sig_{\color{gray}35, e, 1}$ & \cellcolor{neg!20}$-3.8\sig_{\color{gray}31, e, 1}$ & \cellcolor{neg!22}$-4.3\sig_{\color{gray}30, r, 1}$ & \cellcolor{neg!23}$-4.5\sig_{\color{gray}29, r, 1}$ & \cellcolor{pos!46}$+8.9\sig_{\color{gray}51, s, 1}$ & \cellcolor{pos!50}$+10.6\sig_{\color{gray}52, s, 1}$ \\
associated & \cellcolor{pos!3}$+0.5\sig_{\color{gray}44, r, 1}$ & \cellcolor{neg!1}$-0.2\sig_{\color{gray}40, r, 1}$ & \cellcolor{pos!33}$+6.4\sig_{\color{gray}50, s, 5}$ & \cellcolor{pos!19}$+3.6\sig_{\color{gray}49, s, 5}$ & \cellcolor{neg!4}$-0.8\sig_{\color{gray}41, r, 1}$ & \cellcolor{pos!4}$+0.7\sig_{\color{gray}40, r, 1}$ & \cellcolor{pos!50}$+11.2\sig_{\color{gray}57, e, 2}$ & \cellcolor{pos!50}$+11.7\sig_{\color{gray}55, e, 2}$ \\
important & \cellcolor{neg!9}$-1.7\sig_{\color{gray}36, r, 1}$ & \cellcolor{neg!14}$-2.7\sig_{\color{gray}32, r, 1}$ & \cellcolor{neg!27}$-5.2\sig_{\color{gray}26, e, 1}$ & \cellcolor{neg!19}$-3.7\sig_{\color{gray}30, e, 1}$ & \cellcolor{pos!4}$+0.8\sig_{\color{gray}43, e, 5}$ & \cellcolor{pos!24}$+4.6\sig_{\color{gray}52, e, 5}$ & \cellcolor{pos!50}$+42.3\sig_{\color{gray}72, r, 5}$ & \cellcolor{pos!50}$+49.9\sig_{\color{gray}73, r, 5}$ \\
\midrule
\multicolumn{9}{l}{Sub-Words}\\
\midrule
relevancy & \cellcolor{pos!4}$+0.7\sig_{\color{gray}42, e, 5}$ & \cellcolor{pos!11}$+2.1\sig_{\color{gray}42, e, 5}$ & \cellcolor{pos!50}$+12.9\sig_{\color{gray}54, s, 5}$ & \cellcolor{pos!50}$+17.6\sig_{\color{gray}57, s, 5}$ & \cellcolor{neg!20}$-3.8\sig_{\color{gray}34, r, 1}$ & \cellcolor{neg!18}$-3.4\sig_{\color{gray}34, r, 1}$ & \cellcolor{neg!32}$-6.2\sig_{\color{gray}41, r, 5}$ & \cellcolor{neg!7}$-1.4\sig_{\color{gray}44, r, 5}$ \\
relevance & \cellcolor{neg!10}$-1.9\sig_{\color{gray}42, e, 5}$ & \cellcolor{neg!19}$-3.7\sig_{\color{gray}36, e, 5}$ & \cellcolor{neg!12}$-2.3\sig_{\color{gray}44, s, 5}$ & \cellcolor{pos!8}$+1.5\sig_{\color{gray}44, s, 5}$ & \cellcolor{pos!26}$+4.9\sig_{\color{gray}49, s, 5}$ & \cellcolor{pos!50}$+13.4\sig_{\color{gray}52, s, 5}$ & \cellcolor{neg!45}$-8.6\sig_{\color{gray}31, r, 1}$ & \cellcolor{neg!26}$-5.0\sig_{\color{gray}40, r, 1}$ \\
relevantly & \cellcolor{pos!7}$+1.3\sig_{\color{gray}49, r, 1}$ & \cellcolor{pos!10}$+2.0\sig_{\color{gray}49, r, 1}$ & \cellcolor{pos!50}$+13.5\sig_{\color{gray}61, s, 5}$ & \cellcolor{pos!50}$+14.1\sig_{\color{gray}61, s, 5}$ & \cellcolor{neg!1}$-0.2\sig_{\color{gray}40, r, 1}$ & \cellcolor{pos!8}$+1.5\sig_{\color{gray}47, r, 1}$ & \cellcolor{neg!47}$-9.0\sig_{\color{gray}29, r, 1}$ & \cellcolor{neg!33}$-6.3\sig_{\color{gray}35, r, 1}$ \\
irrelevant & \cellcolor{neg!7}$-1.4\sig_{\color{gray}34, e, 1}$ & \cellcolor{pos!6}$+1.2\sig_{\color{gray}35, e, 1}$ & \cellcolor{pos!50}$+30.5\sig_{\color{gray}68, s, 5}$ & \cellcolor{pos!50}$+34.5\sig_{\color{gray}69, s, 5}$ & \cellcolor{neg!20}$-3.8\sig_{\color{gray}34, r, 1}$ & \cellcolor{pos!1}$+0.2\sig_{\color{gray}45, r, 1}$ & \cellcolor{neg!37}$-7.1\sig_{\color{gray}31, r, 1}$ & \cellcolor{neg!5}$-1.0\sig_{\color{gray}38, r, 1}$ \\
\bottomrule
\end{tabular}}
\label{tab:scale}
\end{table*}

\begin{table*}[t]
\centering
\setlength{\tabcolsep}{4pt}
\caption{Efficacy of paraphrasing (Par.) and prepending a summary (Sum.) to rank 100 on various sizes of monoT5 in terms of MRC and success rate (grey subscript).  Significant results are denoted with $^*$ (Students t-test p < 0.05).}
\begin{tabular}{@{}llrrrrrrrr@{}}
\toprule
& & \multicolumn{2}{c}{monoT5$_\text{small}$} & \multicolumn{2}{c}{monoT5$_\text{base}$} & \multicolumn{2}{c}{monoT5$_\text{large}$} & \multicolumn{2}{c}{monoT5$_\text{3B}$}\\

\cmidrule(r{3pt}){3-4}
\cmidrule(r{3pt}){5-6}
\cmidrule(r{3pt}){7-8}
\cmidrule(r{3pt}){9-10}

& LLM & \multicolumn{1}{c}{DL19} & \multicolumn{1}{c}{DL20} & \multicolumn{1}{c}{DL19} & \multicolumn{1}{c}{DL20} & \multicolumn{1}{c}{DL19} & \multicolumn{1}{c}{DL20} & \multicolumn{1}{c}{DL19} & \multicolumn{1}{c}{DL20}  \\
\midrule
\multirow{2}{*}{\rotatebox[origin=c]{90}{\parbox[c]{2.5em}{\centering \textbf{Par.}}}}& Alpaca  & \cellcolor{pos!42}$+2.7\sig_{\color{gray}52}$ & \cellcolor{pos!39}$+2.6\sig_{\color{gray}53}$ & \cellcolor{pos!37}$+2.4\sig_{\color{gray}51}$ & \cellcolor{pos!30}$+1.9\sig_{\color{gray}50}$ & \cellcolor{pos!23}$+1.5\sig_{\color{gray}46}$ & \cellcolor{pos!27}$+1.7\sig_{\color{gray}46}$ & \cellcolor{pos!22}$+1.4\sig_{\color{gray}46}$ & \cellcolor{pos!16}$+1.0_{\color{gray}44}$ \\
& ChatGPT & \cellcolor{pos!27}$+1.7\sig_{\color{gray}52}$ & \cellcolor{pos!16}$+1.0_{\color{gray}50}$ & \cellcolor{pos!47}$+3.0\sig_{\color{gray}56}$ & \cellcolor{pos!34}$+2.2\sig_{\color{gray}54}$ & \cellcolor{pos!19}$+1.2_{\color{gray}50}$ & \cellcolor{pos!10}$+0.6_{\color{gray}48}$ & \cellcolor{pos!9}$+0.6_{\color{gray}46}$ & \cellcolor{neg!2}$-0.1_{\color{gray}46}$ \\
\midrule
\multirow{2}{*}{\rotatebox[origin=c]{90}{\parbox[c]{2.5em}{\centering \textbf{Sum.}}}}& Alpaca & \cellcolor{pos!34}$+2.2\sig_{\color{gray}47}$ & \cellcolor{pos!32}$+2.1\sig_{\color{gray}48}$ & \cellcolor{pos!45}$+2.9\sig_{\color{gray}53}$ & \cellcolor{pos!39}$+2.5\sig_{\color{gray}51}$ & \cellcolor{pos!33}$+2.2\sig_{\color{gray}49}$ & \cellcolor{pos!35}$+2.3\sig_{\color{gray}49}$ & \cellcolor{pos!50}$+3.3\sig_{\color{gray}55}$ & \cellcolor{pos!43}$+2.8\sig_{\color{gray}54}$ \\
& ChatGPT & \cellcolor{pos!23}$+1.5\sig_{\color{gray}47}$ & \cellcolor{pos!17}$+1.1\sig_{\color{gray}47}$ & \cellcolor{pos!29}$+1.9\sig_{\color{gray}50}$ & \cellcolor{pos!10}$+0.6_{\color{gray}46}$ & \cellcolor{pos!10}$+0.6_{\color{gray}45}$ & \cellcolor{pos!16}$+1.0_{\color{gray}45}$ & \cellcolor{pos!15}$+1.0_{\color{gray}47}$ & \cellcolor{pos!6}$+0.4_{\color{gray}45}$ \\
\bottomrule
\end{tabular}
\label{tab:scale-prompt}
\end{table*}

\begin{figure*}[t]
    \vspace*{-.7cm}
    \centering
    \subfloat[][Attack on monoT5 Variants]{\includegraphics[width=0.5\textwidth]{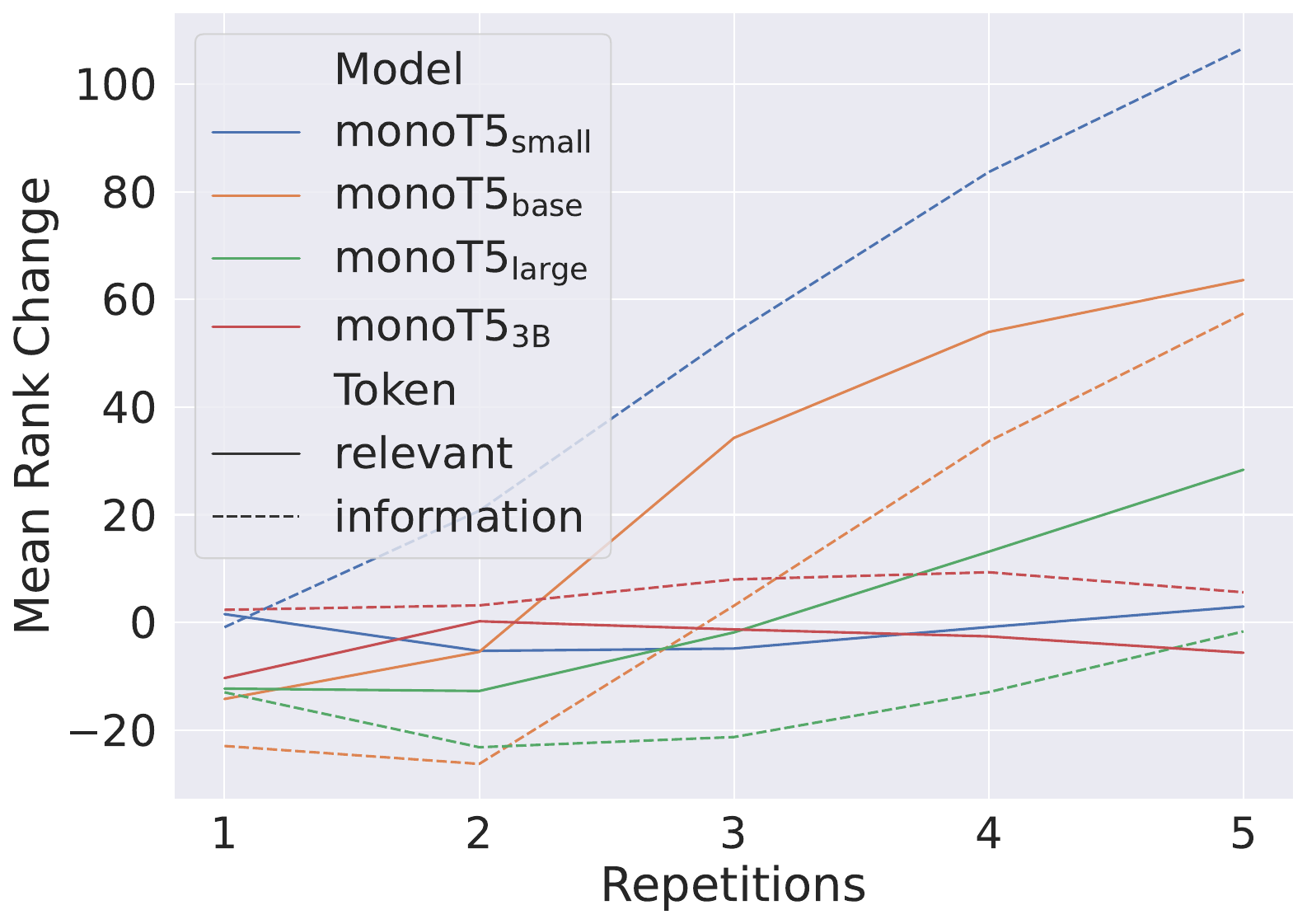}
    }
    \subfloat[][Attack on Multiple NRMs]{\includegraphics[width=0.5\textwidth]{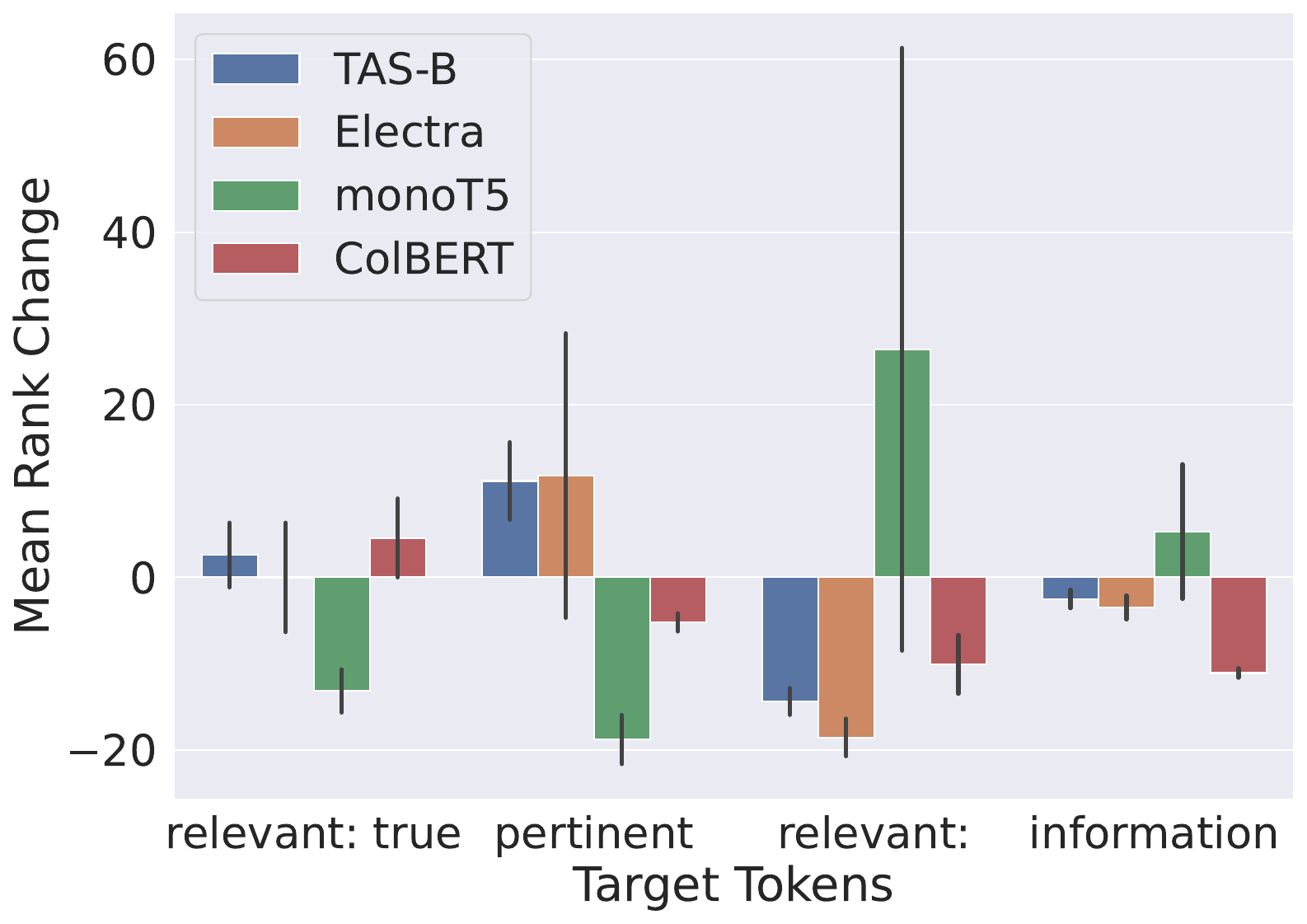}
    }
    \caption{An overview of (a) the scaling of rank improvement for the number of token repetitions of control and prompt tokens with maximum MRC and (b) the variance of repetitions on different neural models for strongest settings.}
    \label{fig:mrc}
\end{figure*}

\paragraph{Transfer of Injection Attacks.} Table \ref{tab:transfer} shows that though monoT5 is most affected by the injection of prompt tokens as illustrated in Figure \ref{fig:mrc}(b), generalisation across neural models can be seen in cases of the injection containing the token `relevant' beyond the constraints of our attack model outlined in Section \ref{sec:attack-model}. Due to BM25 penalties for document length and the addition of tokens that are almost guaranteed not to be contained in the evaluation queries, the attack fails to improve document rank across all token groups\footnote{In the cases of `information' and `related', these words are present within the default PyTerrier stop-words list and as such the document becomes duplicated causing to a tie break, this leads to the small rank change with 0.0\% success rate}. 

Control tokens can greatly influence the monoT5 ranking of documents. However, they do not generalize beyond T5, suggesting that the prompt structure from which these controls were inspired has a significant ranking impact for prompt-based relevance models. We observe mixed results when injecting synonyms for `relevant.' `Significant' is effective, improving document scores for both cross-encoders. TAS-B is also affected. However, ColBERT is unaffected and insensitive to synonyms due to its max pooling token-level similarity computation that may ignore injected tokens, contrasting the deeper interaction of cross-encoders or the passage-level similarity of standard bi-encoders.

Subwords significantly improve the MRC for both cross-encoders, indicating that injecting words containing the token `relevant' has a positive impact. Hence, filtering keyword-stuffing attacks on neural models may be more challenging as tokenization allows hiding attack tokens that lexical models ignore.

Successful attacks on neural models frequently involve multiple repetitions of injected tokens (as can be observed as the 3$^{\text{rd}}$ subscript of each attack in Table \ref{tab:transfer}). This response is unexpectedly similar to a lexical model but does not depend on the frequency of query terms and remains context-agnostic (e.g., the upward trend in Figure~\ref{fig:mrc}(a)). We also observe that BERT-based architectures generally prefer the injection of tokens to the end in contrast with monoT5, which almost always prefers injection at the start (as can be observed as the 2$^{\text{nd}}$ subscript of each attack in Table \ref{tab:transfer}). The stronger generalization of appended injections may suggest that it is a more effective attack when unsure of the language model used in the targeted relevance model.

\paragraph{Transfer of LLM Re-Writing Attacks.} Table~\ref{tab:transfer-prompt} shows that cross-encoders are weaker in paraphrasing and summary attacks. In all cases, a significant improvement is found; however, summaries from ChatGPT fail to improve rank over 50\% of the time in monoT5 on DL20. Rank improvements when attacking TAS-B with paraphrasing and summary are small, only improving rank in over 50\% of documents on DL19. ColBERT is generally not affected by a summary reflecting a general in-variance to our attacks (further outlined by low variance observed in Figure \ref{fig:mrc}(b)). Document rank significantly drops in all paraphrasing attacks against BM25; this can be attributed to the increase in document length from adding the tokens `relevant' and `true' as well as the potential for an LLM re-write to re-phrase terms, which may cause lexical mismatch. However, BM25 rank is improved by the injection of a summary showing that both LLMs have captured useful terms in their summary; as this attack also transfers to cross-encoders, it is an effective attack against a larger search pipeline.
 
\begin{table*}[t]
\footnotesize
\setlength{\tabcolsep}{1pt}
\caption{The MRC and SR (grey subscript) of keyword stuffing on neural models. Significant changes denoted by $^*$ (Bonferroni corrected t-test at p < 0.05).}
\adjustbox{width=\textwidth}{
\begin{tabular}{@{}lcccccccccc@{}}
\toprule
& \multicolumn{2}{c}{BM25} & \multicolumn{2}{c}{ColBERT} & \multicolumn{2}{c}{TAS-B} & \multicolumn{2}{c}{monoT5} & \multicolumn{2}{c}{Electra}\\

\cmidrule(r{3pt}){2-3}
\cmidrule(r{3pt}){4-5}
\cmidrule(r{3pt}){6-7}
\cmidrule(r{3pt}){8-9}
\cmidrule(r{3pt}){10-11}

Token & \multicolumn{1}{c}{DL19} & \multicolumn{1}{c}{DL20} & \multicolumn{1}{c}{DL19} & \multicolumn{1}{c}{DL20} & \multicolumn{1}{c}{DL19} & \multicolumn{1}{c}{DL20} & \multicolumn{1}{c}{DL19} & \multicolumn{1}{c}{DL20} & \multicolumn{1}{c}{DL19} & \multicolumn{1}{c}{DL20}\\

\midrule
\multicolumn{6}{l}{Prompt Tokens} \\
\midrule
true & \cellcolor{neg!50}$-22.0\sig_{\color{gray}0, s, 1}$ & \cellcolor{neg!50}$-22.7\sig_{\color{gray}0, s, 1}$ & \cellcolor{pos!12}$+2.4\sig_{\color{gray}36, s, 1}$ & \cellcolor{pos!16}$+3.2\sig_{\color{gray}34, s, 1}$ & \cellcolor{neg!2}$-0.3\sig_{\color{gray}42, r, 1}$ & \cellcolor{neg!2}$-0.5\sig_{\color{gray}35, r, 1}$ & \cellcolor{neg!46}$-9.1\sig_{\color{gray}22, r, 1}$ & \cellcolor{neg!48}$-9.4\sig_{\color{gray}21, r, 1}$ & \cellcolor{pos!6}$+1.2\sig_{\color{gray}46, e, 5}$ & \cellcolor{pos!16}$+3.2\sig_{\color{gray}47, e, 5}$ \\
false & \cellcolor{neg!50}$-22.0\sig_{\color{gray}0, s, 1}$ & \cellcolor{neg!50}$-22.7\sig_{\color{gray}0, s, 1}$ & \cellcolor{neg!30}$-6.0\sig_{\color{gray}16, e, 1}$ & \cellcolor{pos!21}$+4.1\sig_{\color{gray}38, e, 1}$ & \cellcolor{neg!24}$-4.8\sig_{\color{gray}30, e, 1}$ & \cellcolor{pos!3}$+0.6\sig_{\color{gray}46, e, 1}$ & \cellcolor{neg!4}$-0.8\sig_{\color{gray}46, s, 5}$ & \cellcolor{neg!14}$-2.7\sig_{\color{gray}33, s, 5}$ & \cellcolor{neg!5}$-1.1\sig_{\color{gray}43, e, 5}$ & \cellcolor{pos!13}$+2.6\sig_{\color{gray}45, e, 5}$ \\
relevant: & \cellcolor{neg!50}$-22.0\sig_{\color{gray}0, e, 1}$ & \cellcolor{neg!50}$-22.7\sig_{\color{gray}0, e, 1}$ & \cellcolor{pos!27}$+5.3\sig_{\color{gray}49, e, 5}$ & \cellcolor{pos!8}$+1.6\sig_{\color{gray}45, e, 5}$ & \cellcolor{pos!24}$+4.7\sig_{\color{gray}50, e, 5}$ & \cellcolor{pos!7}$+1.3\sig_{\color{gray}45, e, 5}$ & \cellcolor{pos!50}$+63.6\sig_{\color{gray}78, s, 5}$ & \cellcolor{pos!50}$+51.2\sig_{\color{gray}75, s, 5}$ & \cellcolor{pos!35}$+6.9\sig_{\color{gray}56, e, 5}$ & \cellcolor{pos!27}$+5.4\sig_{\color{gray}51, e, 5}$ \\
relevant: true & \cellcolor{neg!50}$-41.1\sig_{\color{gray}0, s, 1}$ & \cellcolor{neg!50}$-42.9\sig_{\color{gray}0, s, 1}$ & \cellcolor{pos!50}$+9.9\sig_{\color{gray}52, e, 5}$ & \cellcolor{pos!17}$+3.3\sig_{\color{gray}44, e, 5}$ & \cellcolor{pos!34}$+6.8\sig_{\color{gray}54, e, 5}$ & \cellcolor{neg!10}$-2.0\sig_{\color{gray}39, e, 5}$ & \cellcolor{pos!50}$+31.1\sig_{\color{gray}64, s, 5}$ & \cellcolor{pos!50}$+18.3\sig_{\color{gray}57, s, 5}$ & \cellcolor{pos!24}$+4.7\sig_{\color{gray}52, e, 3}$ & \cellcolor{pos!19}$+3.8\sig_{\color{gray}48, e, 3}$ \\
relevant: false & \cellcolor{neg!50}$-41.1\sig_{\color{gray}0, e, 1}$ & \cellcolor{neg!50}$-42.9\sig_{\color{gray}0, e, 1}$ & \cellcolor{pos!35}$+6.8\sig_{\color{gray}52, e, 5}$ & \cellcolor{pos!35}$+6.9\sig_{\color{gray}51, e, 5}$ & \cellcolor{pos!48}$+9.6\sig_{\color{gray}55, e, 5}$ & \cellcolor{pos!50}$+10.4\sig_{\color{gray}52, e, 5}$ & \cellcolor{pos!50}$+47.4\sig_{\color{gray}71, s, 5}$ & \cellcolor{pos!50}$+32.0\sig_{\color{gray}64, s, 5}$ & \cellcolor{pos!16}$+3.2\sig_{\color{gray}47, e, 5}$ & \cellcolor{pos!17}$+3.4\sig_{\color{gray}43, e, 5}$ \\
\midrule
\multicolumn{6}{l}{Control Tokens}\\
\midrule
bar & \cellcolor{neg!50}$-22.0\sig_{\color{gray}0, s, 1}$ & \cellcolor{neg!50}$-22.7\sig_{\color{gray}0, s, 1}$ & \cellcolor{neg!41}$-8.1\sig_{\color{gray}12, e, 1}$ & \cellcolor{neg!47}$-9.2\sig_{\color{gray}9, e, 1}$ & \cellcolor{neg!15}$-3.0\sig_{\color{gray}44, r, 1}$ & \cellcolor{neg!23}$-4.5\sig_{\color{gray}38, r, 1}$ & \cellcolor{neg!18}$-3.5\sig_{\color{gray}36, e, 2}$ & \cellcolor{pos!3}$+0.6\sig_{\color{gray}41, e, 2}$ & \cellcolor{neg!36}$-7.2\sig_{\color{gray}25, r, 1}$ & \cellcolor{neg!37}$-7.3\sig_{\color{gray}27, r, 1}$ \\
baz & \cellcolor{neg!50}$-22.0\sig_{\color{gray}0, r, 1}$ & \cellcolor{neg!50}$-22.7\sig_{\color{gray}0, r, 1}$ & \cellcolor{neg!4}$-0.8\sig_{\color{gray}18, e, 1}$ & \cellcolor{pos!4}$+0.8\sig_{\color{gray}20, e, 1}$ & \cellcolor{neg!50}$-10.5\sig_{\color{gray}32, e, 1}$ & \cellcolor{pos!10}$+2.0\sig_{\color{gray}48, e, 1}$ & \cellcolor{pos!33}$+6.6\sig_{\color{gray}53, s, 5}$ & \cellcolor{pos!50}$+17.2\sig_{\color{gray}60, s, 5}$ & \cellcolor{pos!7}$+1.4\sig_{\color{gray}44, r, 2}$ & \cellcolor{pos!50}$+10.7\sig_{\color{gray}49, r, 2}$ \\
information: & \cellcolor{neg!12}$-2.4\sig_{\color{gray}0, s, 1}$ & \cellcolor{neg!12}$-2.4\sig_{\color{gray}0, s, 1}$ & \cellcolor{neg!50}$-10.3\sig_{\color{gray}11, r, 1}$ & \cellcolor{neg!50}$-9.8\sig_{\color{gray}3, r, 1}$ & \cellcolor{neg!6}$-1.1\sig_{\color{gray}44, e, 5}$ & \cellcolor{pos!11}$+2.1\sig_{\color{gray}48, e, 5}$ & \cellcolor{pos!50}$+57.4\sig_{\color{gray}77, s, 5}$ & \cellcolor{pos!50}$+41.3\sig_{\color{gray}70, s, 5}$ & \cellcolor{neg!11}$-2.1\sig_{\color{gray}41, e, 5}$ & \cellcolor{neg!1}$-0.2\sig_{\color{gray}40, e, 5}$ \\
information: bar & \cellcolor{neg!50}$-22.0\sig_{\color{gray}0, e, 1}$ & \cellcolor{neg!50}$-22.7\sig_{\color{gray}0, e, 1}$ & \cellcolor{neg!50}$-12.3\sig_{\color{gray}7, e, 1}$ & \cellcolor{neg!50}$-12.7\sig_{\color{gray}4, e, 1}$ & \cellcolor{neg!18}$-3.5\sig_{\color{gray}41, e, 1}$ & \cellcolor{neg!19}$-3.8\sig_{\color{gray}35, e, 1}$ & \cellcolor{pos!50}$+31.6\sig_{\color{gray}70, e, 5}$ & \cellcolor{pos!50}$+38.2\sig_{\color{gray}71, e, 5}$ & \cellcolor{neg!50}$-15.4\sig_{\color{gray}24, r, 1}$ & \cellcolor{neg!50}$-12.2\sig_{\color{gray}25, r, 1}$ \\
information: baz & \cellcolor{neg!50}$-22.0\sig_{\color{gray}0, s, 1}$ & \cellcolor{neg!50}$-22.7\sig_{\color{gray}0, s, 1}$ & \cellcolor{neg!33}$-6.6\sig_{\color{gray}14, e, 1}$ & \cellcolor{neg!21}$-4.3\sig_{\color{gray}16, e, 1}$ & \cellcolor{neg!50}$-10.7\sig_{\color{gray}34, e, 1}$ & \cellcolor{pos!22}$+4.4\sig_{\color{gray}50, e, 1}$ & \cellcolor{pos!50}$+31.0\sig_{\color{gray}61, s, 5}$ & \cellcolor{pos!50}$+37.0\sig_{\color{gray}61, s, 5}$ & \cellcolor{neg!50}$-10.2\sig_{\color{gray}30, e, 3}$ & \cellcolor{pos!15}$+3.0\sig_{\color{gray}39, e, 3}$ \\
relevant: bar & \cellcolor{neg!50}$-41.1\sig_{\color{gray}0, r, 1}$ & \cellcolor{neg!50}$-42.9\sig_{\color{gray}0, r, 1}$ & \cellcolor{neg!12}$-2.4\sig_{\color{gray}29, e, 5}$ & \cellcolor{neg!36}$-7.2\sig_{\color{gray}11, e, 5}$ & \cellcolor{pos!3}$+0.6\sig_{\color{gray}50, e, 5}$ & \cellcolor{neg!23}$-4.6\sig_{\color{gray}32, e, 5}$ & \cellcolor{pos!50}$+32.0\sig_{\color{gray}62, s, 5}$ & \cellcolor{pos!50}$+33.6\sig_{\color{gray}61, s, 5}$ & \cellcolor{neg!39}$-7.7\sig_{\color{gray}33, e, 5}$ & \cellcolor{neg!47}$-9.3\sig_{\color{gray}31, e, 5}$ \\
information: true & \cellcolor{neg!50}$-22.0\sig_{\color{gray}0, r, 1}$ & \cellcolor{neg!50}$-22.7\sig_{\color{gray}0, r, 1}$ & \cellcolor{pos!27}$+5.4\sig_{\color{gray}45, e, 5}$ & \cellcolor{pos!13}$+2.5\sig_{\color{gray}38, e, 5}$ & \cellcolor{pos!9}$+1.9\sig_{\color{gray}51, e, 5}$ & \cellcolor{pos!25}$+4.9\sig_{\color{gray}48, e, 5}$ & \cellcolor{pos!50}$+28.4\sig_{\color{gray}62, s, 5}$ & \cellcolor{pos!50}$+13.5\sig_{\color{gray}54, s, 5}$ & \cellcolor{pos!7}$+1.3\sig_{\color{gray}47, e, 4}$ & \cellcolor{pos!26}$+5.1\sig_{\color{gray}47, e, 4}$ \\
\midrule
\multicolumn{6}{l}{Synonyms}\\
\midrule
pertinent & \cellcolor{neg!50}$-22.0\sig_{\color{gray}0, e, 1}$ & \cellcolor{neg!50}$-22.7\sig_{\color{gray}0, e, 1}$ & \cellcolor{neg!5}$-1.0\sig_{\color{gray}24, r, 1}$ & \cellcolor{neg!6}$-1.2\sig_{\color{gray}29, r, 1}$ & \cellcolor{pos!50}$+15.4\sig_{\color{gray}56, e, 5}$ & \cellcolor{pos!50}$+14.9\sig_{\color{gray}54, e, 5}$ & \cellcolor{neg!24}$-4.7\sig_{\color{gray}41, s, 5}$ & \cellcolor{neg!3}$-0.7\sig_{\color{gray}44, s, 5}$ & \cellcolor{pos!50}$+30.1\sig_{\color{gray}77, e, 5}$ & \cellcolor{pos!50}$+28.2\sig_{\color{gray}71, e, 5}$ \\
significant & \cellcolor{neg!50}$-22.0\sig_{\color{gray}0, e, 1}$ & \cellcolor{neg!50}$-22.7\sig_{\color{gray}0, e, 1}$ & \cellcolor{pos!10}$+2.0\sig_{\color{gray}42, r, 1}$ & \cellcolor{neg!5}$-1.0\sig_{\color{gray}31, r, 1}$ & \cellcolor{pos!50}$+10.8\sig_{\color{gray}65, e, 5}$ & \cellcolor{pos!50}$+9.9\sig_{\color{gray}59, e, 5}$ & \cellcolor{pos!50}$+11.3\sig_{\color{gray}55, s, 5}$ & \cellcolor{pos!42}$+8.3\sig_{\color{gray}50, s, 5}$ & \cellcolor{pos!50}$+27.1\sig_{\color{gray}75, e, 5}$ & \cellcolor{pos!50}$+29.2\sig_{\color{gray}76, e, 5}$ \\
related & \cellcolor{neg!12}$-2.4\sig_{\color{gray}0, s, 5}$ & \cellcolor{neg!12}$-2.4\sig_{\color{gray}0, s, 5}$ & \cellcolor{neg!10}$-2.0\sig_{\color{gray}24, r, 1}$ & \cellcolor{neg!20}$-3.9\sig_{\color{gray}23, r, 1}$ & \cellcolor{neg!6}$-1.2\sig_{\color{gray}42, r, 1}$ & \cellcolor{neg!18}$-3.6\sig_{\color{gray}35, r, 1}$ & \cellcolor{neg!10}$-2.1\sig_{\color{gray}35, e, 1}$ & \cellcolor{neg!19}$-3.8\sig_{\color{gray}31, e, 1}$ & \cellcolor{neg!19}$-3.7\sig_{\color{gray}32, r, 1}$ & \cellcolor{neg!23}$-4.5\sig_{\color{gray}30, r, 1}$ \\
associated & \cellcolor{neg!50}$-22.0\sig_{\color{gray}0, s, 1}$ & \cellcolor{neg!50}$-22.7\sig_{\color{gray}0, s, 1}$ & \cellcolor{neg!3}$-0.5\sig_{\color{gray}34, r, 1}$ & \cellcolor{neg!2}$-0.4\sig_{\color{gray}32, r, 1}$ & \cellcolor{neg!5}$-0.9\sig_{\color{gray}42, r, 1}$ & \cellcolor{neg!10}$-1.9\sig_{\color{gray}38, r, 1}$ & \cellcolor{pos!32}$+6.4\sig_{\color{gray}50, s, 5}$ & \cellcolor{pos!18}$+3.6\sig_{\color{gray}49, s, 5}$ & \cellcolor{neg!4}$-0.8\sig_{\color{gray}44, r, 3}$ & \cellcolor{neg!9}$-1.8\sig_{\color{gray}40, r, 3}$ \\
important & \cellcolor{neg!50}$-22.0\sig_{\color{gray}0, r, 1}$ & \cellcolor{neg!50}$-22.7\sig_{\color{gray}0, r, 1}$ & \cellcolor{pos!9}$+1.7\sig_{\color{gray}36, r, 1}$ & \cellcolor{neg!20}$-3.9\sig_{\color{gray}19, r, 1}$ & \cellcolor{pos!24}$+4.7\sig_{\color{gray}49, e, 5}$ & \cellcolor{pos!17}$+3.4\sig_{\color{gray}47, e, 5}$ & \cellcolor{neg!26}$-5.2\sig_{\color{gray}26, e, 1}$ & \cellcolor{neg!19}$-3.7\sig_{\color{gray}30, e, 1}$ & \cellcolor{pos!50}$+25.6\sig_{\color{gray}78, e, 5}$ & \cellcolor{pos!50}$+28.3\sig_{\color{gray}79, e, 5}$ \\
\midrule
\multicolumn{6}{l}{Sub-Words}\\
\midrule
relevancy & \cellcolor{neg!50}$-22.0\sig_{\color{gray}0, r, 1}$ & \cellcolor{neg!50}$-22.7\sig_{\color{gray}0, r, 1}$ & \cellcolor{pos!36}$+7.1\sig_{\color{gray}35, s, 1}$ & \cellcolor{pos!38}$+7.6\sig_{\color{gray}38, s, 1}$ & \cellcolor{neg!7}$-1.4\sig_{\color{gray}47, r, 1}$ & \cellcolor{pos!5}$+1.0\sig_{\color{gray}46, r, 1}$ & \cellcolor{pos!50}$+12.9\sig_{\color{gray}54, s, 5}$ & \cellcolor{pos!50}$+17.6\sig_{\color{gray}57, s, 5}$ & \cellcolor{pos!50}$+27.6\sig_{\color{gray}70, e, 5}$ & \cellcolor{pos!50}$+30.9\sig_{\color{gray}68, e, 5}$ \\
relevance & \cellcolor{neg!50}$-22.0\sig_{\color{gray}0, r, 1}$ & \cellcolor{neg!50}$-22.7\sig_{\color{gray}0, r, 1}$ & \cellcolor{neg!14}$-2.7\sig_{\color{gray}28, r, 1}$ & \cellcolor{neg!11}$-2.1\sig_{\color{gray}30, r, 1}$ & \cellcolor{neg!8}$-1.7\sig_{\color{gray}41, r, 1}$ & \cellcolor{neg!6}$-1.1\sig_{\color{gray}40, r, 1}$ & \cellcolor{neg!12}$-2.3\sig_{\color{gray}44, s, 5}$ & \cellcolor{pos!8}$+1.5\sig_{\color{gray}44, s, 5}$ & \cellcolor{neg!11}$-2.3\sig_{\color{gray}45, r, 2}$ & \cellcolor{pos!3}$+0.5\sig_{\color{gray}45, r, 2}$ \\
relevantly & \cellcolor{neg!50}$-22.0\sig_{\color{gray}0, r, 1}$ & \cellcolor{neg!50}$-22.7\sig_{\color{gray}0, r, 1}$ & \cellcolor{pos!2}$+0.4\sig_{\color{gray}32, r, 1}$ & \cellcolor{neg!3}$-0.6\sig_{\color{gray}31, r, 1}$ & \cellcolor{pos!31}$+6.1\sig_{\color{gray}57, e, 5}$ & \cellcolor{pos!30}$+5.9\sig_{\color{gray}55, e, 5}$ & \cellcolor{pos!50}$+13.5\sig_{\color{gray}61, s, 5}$ & \cellcolor{pos!50}$+14.1\sig_{\color{gray}61, s, 5}$ & \cellcolor{pos!50}$+22.5\sig_{\color{gray}66, r, 5}$ & \cellcolor{pos!50}$+27.0\sig_{\color{gray}65, r, 5}$ \\
irrelevant & \cellcolor{neg!50}$-22.0\sig_{\color{gray}0, e, 1}$ & \cellcolor{neg!50}$-22.7\sig_{\color{gray}0, e, 1}$ & \cellcolor{pos!13}$+2.6\sig_{\color{gray}42, r, 1}$ & \cellcolor{pos!4}$+0.8\sig_{\color{gray}38, r, 1}$ & \cellcolor{pos!30}$+5.9\sig_{\color{gray}53, r, 1}$ & \cellcolor{pos!21}$+4.1\sig_{\color{gray}48, r, 1}$ & \cellcolor{pos!50}$+30.5\sig_{\color{gray}68, s, 5}$ & \cellcolor{pos!50}$+34.5\sig_{\color{gray}69, s, 5}$ & \cellcolor{pos!50}$+11.5\sig_{\color{gray}60, e, 5}$ & \cellcolor{pos!50}$+15.1\sig_{\color{gray}60, e, 5}$ \\
\bottomrule
\end{tabular}}
\label{tab:transfer}
\end{table*}

\begin{table*}
\footnotesize
\setlength{\tabcolsep}{1pt}
\caption{Overview of the MRC and SR (subscript) for re-writing with paraphrasing (Par.) and by prepending a summary (Sum.) for Alpaca and ChatGPT. Significant changes denoted with $^*$ (Bonferroni corrected t-test at p < 0.05).}
    \adjustbox{width=\textwidth}{

\begin{tabular*}{\textwidth}{@{}llrrrrrrrrrrrr@{}}
\toprule
& & \multicolumn{2}{c}{BM25} & \multicolumn{2}{c}{ColBERT} & \multicolumn{2}{c}{TAS-B} & \multicolumn{2}{c}{monoT5} & \multicolumn{2}{c}{Electra}\\

\cmidrule(r{3pt}){3-4}
\cmidrule(r{3pt}){5-6}
\cmidrule(r{3pt}){7-8}
\cmidrule(r{3pt}){9-10}
\cmidrule{11-12}

& LLM & \multicolumn{1}{c}{DL19} & \multicolumn{1}{c}{DL20} & \multicolumn{1}{c}{DL19} & \multicolumn{1}{c}{DL20} & \multicolumn{1}{c}{DL19} & \multicolumn{1}{c}{DL20} & \multicolumn{1}{c}{DL19} & \multicolumn{1}{c}{DL20} & \multicolumn{1}{c}{DL19} & \multicolumn{1}{c}{DL20}\\

\midrule
\multirow{2}{*}{\rotatebox[origin=c]{90}{\parbox[c]{2.5em}{\centering \textbf{Par.}}}} & Alpaca & \cellcolor{neg!50}$-14.9\sig_{\color{gray}20}$ & \cellcolor{neg!50}$-13.6\sig_{\color{gray}20}$ & \cellcolor{pos!16}$+1.3\sig_{\color{gray}45}$ & \cellcolor{pos!13}$+1.0_{\color{gray}44}$ & \cellcolor{pos!5}$+0.4_{\color{gray}48}$ & \cellcolor{pos!0}\phantom{+}$0.0_{\color{gray}46}$ & \cellcolor{pos!30}$+2.4\sig_{\color{gray}51}$ & \cellcolor{pos!24}$+1.9\sig_{\color{gray}50}$ & \cellcolor{pos!50}$+4.1\sig_{\color{gray}55}$ & \cellcolor{pos!46}$+3.8\sig_{\color{gray}54}$ \\

& ChatGPT & \cellcolor{neg!50}$-27.1\sig_{\color{gray}9\phantom{0}}$ & \cellcolor{neg!50}$-26.9\sig_{\color{gray}9\phantom{0}}$ & \cellcolor{pos!16}$+1.3\sig_{\color{gray}50}$ & \cellcolor{pos!3}$+0.2_{\color{gray}48}$ & \cellcolor{pos!16}$+1.3\sig_{\color{gray}52}$ & \cellcolor{pos!6}$+0.5_{\color{gray}48}$ & \cellcolor{pos!37}$+3.0\sig_{\color{gray}56}$ & \cellcolor{pos!27}$+2.2\sig_{\color{gray}54}$ & \cellcolor{pos!32}$+2.6\sig_{\color{gray}55}$ & \cellcolor{pos!23}$+1.9\sig_{\color{gray}53}$ \\

\midrule
\multirow{2}{*}{\rotatebox[origin=c]{90}{\parbox[c]{2.5em}{\centering \textbf{Sum.}}}} & Alpaca & \cellcolor{pos!48}$+\phantom{0}3.9\sig_{\color{gray}56}$ & \cellcolor{pos!48}$+\phantom{0}3.9\sig_{\color{gray}56}$ & \cellcolor{pos!0}\phantom{+}$0.0_{\color{gray}40}$ & \cellcolor{neg!2}$-0.2_{\color{gray}38}$ & \cellcolor{pos!21}$+1.7\sig_{\color{gray}48}$ & \cellcolor{pos!16}$+1.3\sig_{\color{gray}47}$ & \cellcolor{pos!36}$+2.9\sig_{\color{gray}53}$ & \cellcolor{pos!31}$+2.5\sig_{\color{gray}51}$ & \cellcolor{pos!49}$+4.0\sig_{\color{gray}54}$ & \cellcolor{pos!39}$+3.2\sig_{\color{gray}53}$ \\

& ChatGPT & \cellcolor{pos!37}$+\phantom{0}3.0\sig_{\color{gray}55}$ & \cellcolor{pos!30}$+\phantom{0}2.4\sig_{\color{gray}51}$ & \cellcolor{neg!24}$-2.0\sig_{\color{gray}35}$ & \cellcolor{neg!22}$-1.8\sig_{\color{gray}34}$ & \cellcolor{pos!2}$+0.1_{\color{gray}45}$ & \cellcolor{neg!2}$-0.2_{\color{gray}42}$ & \cellcolor{pos!23}$+1.9\sig_{\color{gray}50}$ & \cellcolor{pos!8}$+0.6_{\color{gray}46}$ & \cellcolor{pos!37}$+3.0\sig_{\color{gray}54}$ & \cellcolor{pos!29}$+2.4\sig_{\color{gray}52}$ \\
\bottomrule
\end{tabular*}}
\label{tab:transfer-prompt}
\end{table*}

\subsection{A Search Provider's Perspective}
\label{sec:search-provider-perspective}

\begin{table*}[tb]
\setlength{\tabcolsep}{1.7pt}
\renewcommand{\arraystretch}{1.}
\caption{The retrieval effectiveness when adversarial attacks are applied to non-relevant documents (worst case), to no documents (original case), or to only relevant documents (best case). We report nDCG@10 and Precision@10 where $^{*}$ marks Bonferroni corrected significant changes to the no-attack scenario.}
\footnotesize
\label{table-retrieval-effectiveness}
\begin{tabular*}{\textwidth}{@{}lcccccc@{\quad}cccccc@{}}

    \toprule
    
    & \multicolumn{6}{@{}c@{}}{\textbf{TREC DL 19}} & \multicolumn{6}{@{}c@{}}{\textbf{TREC DL 20}} \\
    
    \cmidrule(r{20pt}){2-7} \cmidrule(){8-13}
    & \multicolumn{3}{@{}c@{}}{\textbf{nDCG@10}} & \multicolumn{3}{@{}c@{}}{\textbf{Precision@10\phantom{mm}}} & \multicolumn{3}{@{}c@{}}{\textbf{nDCG@10}} & \multicolumn{3}{@{}c@{}}{\textbf{Precision@10}} \\

    \cmidrule(r{4pt}){2-4} \cmidrule(r{20pt}){5-7} \cmidrule(r{4pt}){8-10} \cmidrule{11-13} 

    & Worst & Ori.\ & Best & Worst & Ori.\ & Best & Worst & Ori.\ & Best & Worst & Ori.\ & Best \\
    \midrule

    BM25 & 0.48$\phantom{^{*}}$ & 0.48$\phantom{^{*}}$ & 0.48$\phantom{^{*}}$ & 0.60$\phantom{^{*}}$ & 0.60$\phantom{^{*}}$ & 0.60$\phantom{^{*}}$ & 0.49$\phantom{^{*}}$ & 0.49$\phantom{^{*}}$ & 0.49$\phantom{^{*}}$ & 0.58$\phantom{^{*}}$ & 0.58$\phantom{^{*}}$ & 0.58$\phantom{^{*}}$ \\
    ColBERT & 0.66$\phantom{^{*}}$ & 0.68$\phantom{^{*}}$ & 0.71$^{*}$ & 0.74$^{*}$ & 0.77$\phantom{^{*}}$ & 0.82$^{*}$ & 0.62$^{*}$ & 0.66$\phantom{^{*}}$ & 0.69$^{*}$ & 0.64$^{*}$ & 0.69$\phantom{^{*}}$ & 0.73$^{*}$ \\
    Electra & 0.69$^{*}$ & 0.71$\phantom{^{*}}$ & 0.73$^{*}$ & 0.77$^{*}$ & 0.80$\phantom{^{*}}$ & 0.83$^{*}$ & 0.67$^{*}$ & 0.70$\phantom{^{*}}$ & 0.73$^{*}$ & 0.70$^{*}$ & 0.74$\phantom{^{*}}$ & 0.78$^{*}$ \\
    monoT5 & 0.67$^{*}$ & 0.70$\phantom{^{*}}$ & 0.73$^{*}$ & 0.74$^{*}$ & 0.79$\phantom{^{*}}$ & 0.85$^{*}$ & 0.64$^{*}$ & 0.68$\phantom{^{*}}$ & 0.72$^{*}$ & 0.66$^{*}$ & 0.71$\phantom{^{*}}$ & 0.77$^{*}$ \\
    TAS-B & 0.67$^{*}$ & 0.69$\phantom{^{*}}$ & 0.72$^{*}$ & 0.75$^{*}$ & 0.78$\phantom{^{*}}$ & 0.82$^{*}$ & 0.62$^{*}$ & 0.66$\phantom{^{*}}$ & 0.70$^{*}$ & 0.68$^{*}$ & 0.71$\phantom{^{*}}$ & 0.76$^{*}$ \\

    \bottomrule
\end{tabular*}
\end{table*}

We assess the impact of our adversarial attacks on the retrieval effectiveness of all models by contrasting hypothetical lower/upper bounds that we obtain in an oracle scenario. For the lower bound, we simulate that only non-relevant documents apply adversarial attacks. We simulate that only relevant documents apply adversarial attacks as an upper bound. In all cases, we select the adversarial attack that causes the highest rank change to report the maximum effect for each document. Following our previous observations that re-writing attacks have a smaller impact than injection attacks, we only include injection attacks in our retrieval effectiveness experiments to maintain our focus on lower/upper bounds. We report nDCG@10 and Precision@10 (albeit controversial~\cite{sakai:2020}, we leave out MRR because MRR has several shortcomings~\cite{fuhr:2017,zobel:2020}). All neural models re-rank the top 1000 documents retrieved by BM25. We report significance compared to the original documents using a Student's t-test with Bonferroni correction.

Table~\ref{table-retrieval-effectiveness} shows the maximum impact of our adversarial attacks by contrasting the worst case (lower bound effectiveness) with the original effectiveness (documents are not manipulated) and the best case (upper bound effectiveness). The injection attacks do not impact BM25, as the retrieval scores never increase by adding non-query tokens to documents. In all other cases, adversarial attacks have a substantial impact on the retrieval effectiveness as the lower and upper bounds introduce, in almost all cases, significant changes, causing our attacks to degrade retrieval effectiveness at scale (the lower bound on nDCG@10 for ColBERT of~0.66 being the only exception). Adversarial attacks have the highest impact on monoT5 (only TAS-B on TREC DL~2020 has the same lower/upper-bound variance of nDCG@10). Importantly, for system-oriented evaluations, we observe that the system rankings are unstable across the different scenarios for nDCG@10 and Precision@10. For instance, monoT5 is with an nDCG@10 of~0.70 more effective than TAS-B with 0.69~on TREC DL~2019 in the original case but less effective in the best case (0.73~for monoT5 vs. 0.72~for TAS-B). Overall, adversarial attacks have a high impact in the comparison, e.g., with the paradigm change introduced by BERT, effectiveness shot up by around 0.08~MRR on the MS~MARCO test set~\cite{lin:2021}, but adversarial attacks introduce even larger changes, e.g., 0.08~nDCG@10 or even 0.11~Precision@10 for monoT5 on TREC DL~2020.

\section{Conclusion}
\label{conclusion}

We presented query-independent adversarial attacks against prompt-based sequence-to-sequence relevance models. By exploiting monoT5's prompt structure, we found the attacks successful in more than~78\%. Furthermore, we showed that these attacks transfer to other classes of relevance models, such as encoder-only cross-encoders and bi-encoders. From a content provider's perspective, these attacks can be seen as an effective SEO approach, resulting in mean rank improvements of over 63 places. From a search provider's perspective, the attacks pose a marked risk to search engine effectiveness, which is an important finding given that the research field of information retrieval is moving towards more prompt-based models. Looking at how to harden neural relevance models against our simple adversarial attacks is an important direction for future work, especially given recent state-of-the-art sequence-to-sequence approaches to ranking and the proposal of automatic data labeling by large language models.

\section*{Acknowledgments}
Partially supported by the European Union's Horizon Europe research and innovation programme under grant agreement No 101070014 (\href{https://doi.org/10.3030/101070014}{OpenWebSearch.EU}).

\begin{raggedright}
\small
\bibliography{ecir24-adversarial-attacks-on-retrieval-models-lit}
\end{raggedright}

\end{document}